\documentclass[10pt,a4paper]{article}

\begin{document}

\title{Uniform rotation\\
{\normalsize To whom it may concern}}
\author{Ll. Bel\thanks{e-mail:  wtpbedil@lg.ehu.es}}

\maketitle

\begin{abstract}
I describe the space-time model of a uniformly rotating frame of reference satisfying the Helmholtz free mobility postulate, as we implemented it in a preceding article, and we discuss the implications of this model as it concerns the problematic of the one way or two ways velocity of light derived from the model and its relationship with the universal constant c.

\end{abstract}

\section{Uniform rotation : the space model}

Starting with Minkowski's line-element using Cartesian coordinates\,\footnote{Reference \cite{Rizzi} contains many contributions to this subject}:

\begin{equation}
\label{1}
ds^2=-dt^2+d{\bar l}^2, \ c=1
\end{equation}
with:

\begin{equation}
\label{2}
d{\bar l}^2=du^2+dv^2+dw^2
\end{equation}
let us consider the time-like congruence $\mathcal{C}$:

\begin{equation}
\label{3}
t=t, \ u=X\cos\,\omega t-Y\sin\,\omega t,
\ v=X\sin\,\omega t+Y\cos\,\omega t, \ z=z.
\end{equation}
defined in the domain:

\begin{equation}
\label{4}
\omega\sqrt{X^2+Y^2+Z^2}< 1.
\end{equation}

The line-element (\ref{1}), using a Weyl-like decomposition, becomes:

\begin{equation}
\label{5}
ds^2=-A^2(-dt+f_XdX+f_YdY)^2+A^{-2}d{\bar s}^2,
\end{equation}
where:

\begin{equation}
\label{6}
A=\sqrt{1-\omega^2 R^2}, \quad R=\sqrt{X^2+Y^2}
\end{equation}

\begin{equation}
\label{7}
f_X=A^{-2}\omega Y ,  \ f_Y=-A^{-2}\omega X
\end{equation}
and:

\begin{equation}
\label{8}
d{\bar s}^2=(1-\omega^2X^2)dX^2+(1-\omega^2Y^2)dY^2+(1-\omega^2R^2)dz^2-2\omega^2XY dXdY.
\end{equation}

To give a new meaning to the space-coordinates the recipe that I proposed in \cite{Bel1}, adapted to the present case, is to introduce a new system of coordinates $x,y,z$ and a flat metric of reference:

\begin{equation}
\label{9}
d{\tilde s}^2=\tilde g_{ij}(x^k)dx^idx^j
\end{equation}
such that:

\begin{equation}
\label{10}
(\bar\Gamma^i_{jk}-\tilde\Gamma^i_{jk})\bar g^{jk}=0,
\end{equation}
$\bar\Gamma$ and $\tilde\Gamma$  being the Christoffel symbols of the 3-dimensional metrics (\ref{8}) and (\ref{9}).
Since these conditions are invariant under adapted coordinate transformations of space:

\begin{equation}
\label{11}
X=X(x,y),\ \ Y=Y(x,y),\ \ z=z,
\end{equation}
we may further require the coordinates $x,y,z$ to  be Cartesian coordinates of (\ref{9}) so that:

\begin{equation}
\label{12}
d{\tilde s}^2=dx^2+dy^2+dz^2
\end{equation}
in which case they will be harmonic coordinates of the space metric (\ref{8}).

The corresponding coordinates transformation is:

\begin{equation}
\label{17}
X=\left(1+\frac14 \omega^2\left(\frac13 x^2+y^2\right)\right)x, \quad Y=\left(1+\frac14 \omega^2\left(\frac13 y^2+x^2\right)\right)y, \quad z=z
\end{equation}
and we get:

\begin{equation}
\label{18}
ds^2=-A^2(-dt+f_xdx+f_ydy)+A^{-2}d\bar s^2
\end{equation}
where now:

\begin{eqnarray}
\nonumber
&& \hspace{-1cm}A=\sqrt{1-\omega^2\rho^2},  \quad \rho =\sqrt{x^2+y^2}\\[2ex]
\label{19}
&& \hspace{-1cm} f_x=\omega y \ \ f_y=-\omega x,
\end{eqnarray}
and:

\begin{equation}
\label{20}
d\bar s^2=\left(1-\frac12\omega^2(x^2-y^2)\right)dx^2
+\left(1-\frac12\omega^2(y^2-x^2)\right)dy^2
+(1-\omega^2\rho^2)dz^2
\end{equation}
where I have neglected the powers of $\omega$ beyond $\omega^2$, as it wil be the case in the sequel of the paper. Notice that the coefficients of $dx^2$ and $dy^2$ might be greater or less than 1.

\section{The speed of light}

The historical measures of the universal constant $c$ were obtained as the quotient $\Delta L/\Delta T$ with separate measures of the distance traveled by a light ray and the time of flight. Or using the formula $\epsilon\mu=c^2$ from separate measures of $\epsilon$, the permitivity of free space, and $\mu$ the permeability of free space; a method in which light and its propagation do not come into play and thus proves that $c$ is much more than the speed of light.
The value of the universal constant $c$ that now has become a definition, since 1983, was measured in 1972 as the product $\lambda\nu$ of the frequency $\nu$ and the wave-length $\lambda$ of a particular cavity resonance \cite{Hall}, and the value that was obtained by no means depended on any relativistic global space-time model nor synchronization protocol. We discuss below light propagation in the framework of the space model presented in the first section and the simplest of the time models where time is represented by the coordinate $t$.

Let us start  with the general line-element (\ref{18}) and the space-metric of reference (\ref{12}). Solving for $dt>0$ the equation $ds^2=0$ we have

\begin{equation}
\label{21}
dt^\pm=\pm f_idx^i+A^{-2}\sqrt{d\bar s}; \quad i=x,y,z
\end{equation}
$dt\pm$, both positives, are therefore the infinitesimal time-elements corresponding to a given direction and its opposite at a given point.

The corresponding speed is:

\begin{equation}
\label{22}
|v^\pm|=\frac{d\tilde s}{dt^\pm}
\end{equation}
where:

\begin{equation}
\label{23}
d\tilde s=\sqrt{dx^2+dy^2+dz^2}
\end{equation}
This shows that in this interpretation the speed of light is anisotropic and time-model dependent\,\footnote{more on that in the next section}.

Using (\ref{19})-(\ref{20}) we get at the approximation that we are considering:

\begin{eqnarray}
&& \hspace{-1cm} dt^\pm=(1+\omega^2 \rho^2)d\tilde s \pm\omega(ydx-xdy) \nonumber \\[2ex]
\label{26}
&& +\frac{\omega^2}{4 d\tilde s}((x^2-y^2)(dy^2-dx^2)-2\rho^2dz^2)
\end{eqnarray}
from where easily follows the one way velocity of light:

\begin{eqnarray}
&& \hspace{-1cm} v^\pm=\frac{d\tilde s}{dt^\pm}=(1-\frac12\omega^2 \rho^2) \mp\omega(y\alpha-x\beta) \nonumber \\[2ex]
\label{v+-}
&& +\frac{\omega^2}{4} ((x^2-y^2)(\beta^2-\alpha^2))-2\omega^2\alpha\beta xy
\end{eqnarray}
where $\alpha,\beta,\gamma$ are the cosines of the direction of propagation.
Introducing the infinitesimal mean $dt$ time that light takes to go from a point $P$ to $P+dP$ and back to $P$, and the mean velocity $v$:

\begin{equation}
\label{v}
dt=\frac12(dt^++dt^-), \quad v=\frac{d\tilde s}{dt}
\end{equation}
$v$ is obtained dropping the term of order $\omega$ of (\ref{v+-}). 

Let us consider, using (\ref{1}), an unconstrained light ray that it is send from a point of coordinates $u_0,v_0,w_0$ at time $0$ so that the parametric equations of the space trajectory are:

\begin{equation}
\label{N1}
u=\alpha t+u_0, \quad v=\beta t+v_0, \quad w=\gamma t+w_0,
\end{equation}

Inverting now (\ref{3}):

\begin{equation}
\label{N2}
t=t, \ X=u\cos\,\omega t+v\sin\,\omega t, \ Y=-u\sin\,\omega t+v\cos\,\omega t, \ z=z.
\end{equation}
as well the coordinate transformation (\ref{17}):

\begin{equation}
\label{N3}
x=\left(1-\frac14 \omega^2\left(\frac13 X^2+Y^2\right)\right)X, \quad y=\left(1-\frac14 \omega^2\left(\frac13 Y^2+X^2\right)\right)Y, \quad z=z,
\end{equation}
we get after the corresponding substitutions and approximation to order $\omega^2$:

\begin{eqnarray}
\label{N4}
&& \hspace{-20mm} x=\alpha t+x_0-(\beta t+y_0)t\omega-\left(\frac12(\alpha t+x_0)t^2-(\frac{1}{12}x_0^2+\frac14 y_0^2)x_0\right)\omega^2 \nonumber \\
&& \hspace{20mm}-\left(\frac{1}{12}(\alpha t+x_0)^2 +\frac14(\beta t+y_0)^2\right)(\alpha t+x_0)\omega^2
\end{eqnarray}

\begin{eqnarray}
\label{N5}
&& \hspace{-20mm} y=\beta t+y_0+\left(\alpha t+x_0)t\omega-(\frac12(\beta t+y_0)t^2-(\frac{1}{12} y_0^2+\frac14 x_0^2)y_0\right)\omega^2 \nonumber \\
&& \hspace{+20mm}-\left(\frac{1}{12}(\beta t+y_0)^2 +\frac14(\alpha t+x_0)^2\right)(\beta t+y_0)\omega^2
\end{eqnarray}
as well as:

\begin{equation}
\label{NN}
z=\gamma t+z_0
\end{equation}
I consider now de case of light constrained to circulate along  a circular ring of radius $r$ with center at the origin. If the circle is in the plane $z=0$ the parametric equations are:

\begin{equation}
\label{N6}
x=r\cos \phi, \quad y=r\sin \phi, \quad z=0
\end{equation}
Using now (\ref{26}) we have:

\begin{equation}
\label{N7}
dt=\left(1-r\omega +\frac14(4(\cos^4\phi-\cos^2\phi)\omega^2r^2)+5r^2\omega^2\right)r d\phi
\end{equation}
and integrating from $0$ to $2\pi$ we we see that the period is

\begin{equation}
\label{N8}
\Delta t_{hor}=2\pi r-2\pi r^2\omega+\frac94\pi r^3\omega^2
\end{equation}

On de other hand if the circle lies in the vertical plane $y=0$ the parametric equations are:

\begin{equation}
\label{N9}
x=r\cos \theta, \quad z=r\sin \theta, \quad y=0
\end{equation}
and from (\ref{26}) we get:

\begin{equation}
\label{N10}
dt=\left(1-\frac14(\cos^2\theta-3)\cos^2\theta\omega^2r^2\right) r d\theta
\end{equation}
that  after integration from $0$ to $2\pi$ yields the value:

\begin{equation}
\label{N11}
\Delta t_{ver}=2\pi r+\frac{9}{16}\pi r^3\omega^2
\end{equation}

\section{Time models}

I have kept the coordinate $t$ unchanged on going from the line-element (\ref{1}) to (\ref{18})and there are good reasons to do it this way as follows from these three considerations: i) $t$ is a cyclic coordinate in both cases; ii) $t$ is still harmonic:
\begin{equation}
\label{41}
\Delta_4 t=0
\end{equation}
where  $\Delta_4$ is the corresponding 4-dimensional Laplace operator, and iii) the time synchronization that $t$ describes is of the chorodesic type,\cite{Bel1}, centered at the axis of rotation. This meaning in particular  that:

\begin{equation}
\label{42}
\left(f_i\right)_0=0, \ \ \left(\frac{\partial^2f_k}{\partial x^i\partial x^j}\right)_0= 0, \quad  i,j,k=x,y,z
\end{equation}
holds on the axis of symmetry.

On the other hand these three properties do not fix the synchronization generated by the time coordinate. Let us consider another time variable defined by:

\begin{equation}
\label{43}
t^\prime=t+\omega\psi(x,y,z)
\end{equation}
so that this new coordinate $t^\prime$ is again a cyclic one.

In this new time model the geometrical elements of the line-element (\ref{}) are modified as follows: $A$ and $d\bar s^2$ remain unchanged but but $f_i$ become:

\begin{equation}
\label{48}
\ f^\prime_i=f_i+\partial_i S,
\end{equation}
This  $f^\prime_i$ transformations very much affect the speed of unconstrained light because of the $f_i$ dependence of $dt$ (\ref{21}) but
do not affect the values of (\ref{N8}) and (\ref{N11}) because for light constrained to a close circuit the contribution of the gradient of $S$ in the line integral calculation is zero.

A general function $S$ would destroy the properties (\ref{41})and (\ref{42}), but this is not the case with the time transformation:
\begin{equation}
\label{46}
t^\prime=t+k\omega(x^2+y^2-2z^2)
\end{equation}
where $k$ is an arbitrary constant. This is important to keep in mind in the analysis of experiments like OPERA \cite{OPERA} for example.

\end{document}